\documentclass{jetpl}
\twocolumn
\lat

\title{Dirac points, spinons and  spin liquid in twisted bilayer graphene}

\rtitle{Dirac points and  spin liquid in twisted bilayer graphene}

\sodtitle{Dirac points, spinons and  spin liquid in twisted bilayer graphene}

\author{V. Yu. Irkhin
\/\thanks{e-mail: valentin.irkhin@imp.uran.ru} and Yu. N. Skryabin}

\rauthor{V. Yu. Irkhin, Yu. N. Skryabin}

\sodauthor{Irkhin, Skryabin}

\address{M. N. Mikheev Institute of Metal Physics, 620108 Ekaterinburg, Russia}

\abstract{
Twisted bilayer graphene is an excellent example of  highly correlated system
demonstrating a nearly flat electron band, the Mott transition and probably a spin liquid state. Besides the one-electron picture, analysis of  Dirac points  is performed  in terms of spinon Fermi surface in the limit of strong correlations.
Application of gauge field theory to describe deconfined spin liquid phase is treated.
Topological quantum transitions, including those from small to large Fermi surface in the presence of van Hove singularities, are discussed.
}


\begin{document}

\maketitle

\textbf{1. Introduction}

Heterostructures of two-dimensional (atomically-thin) materials  attract great attention of scientists owing to their ability to provide novel electronic properties.
Recently,  correlated flat band has been observed in  a graphene bilayer system \cite{moire}.
This band results from the superlattice modulation in the moire structure of two graphene sheets twisted by an angle which is close to the theoretically predicted ``magic angle''.
The temperature dependence of the amplitude of the Shubnikov-de Haas  oscillations demonstrated large electron effective masses and small Fermi velocities.
%two graphene sheets twisted by an angle close to the theoretically predicted 'magic angle' result in a flat band structure near charge neutrality gives rise to a strongly-correlated electronic system. 6

The unique properties of  twisted bilayer graphene (TwBLG) open up a new basis for  many-body quantum phases.
The  accessibility and gate tunability of the flat bands through twist angle may provide the way to a number of exotic correlated systems, including unusual superconductors and quantum spin liquids.
In particular,  for carrier concentration near half of the
superlattice density, $n=\pm n_s/2$ (which corresponds to  half-filling in the effective Hubbard model) the flat band strongly correlated system is Mott-like insulating phase arising from electron localization in the moire superlattice.
The metal-insulator transition at about 4 K is confirmed by measurements of transport properties (conductance)  \cite{moire}. This behavior is qualitatively different from previously reported zero-field insulating behavior  which occurs at  an integer multiple of $\pm n_s$.
% 16 20 and we shall refer to them as half-filling insulating phases (HFIPs).

In a typical Mott insulator, the ground state usually has an
antiferromagnetic spin ordering which is not observed in our system.  Thus we have a Mott paramagnetic ground state that can be described within modern theoretical concepts. This singlet ground state can be treated as a spin liquid.
From this point of view, the TwBLG system is somewhat similar to copper-oxide systems with square lattice, but the situation is even more 	
favorable: suppression of antiferromagnetic ordering owing to frustrations on a triangular lattice is surely justified, unlike cuprates where competition of long-range electron hoppings in the $t-J$ model should be introduced  \cite{Nagaosa}.

In the present paper we discuss possible field-theoretical approaches to describe the TwBLG system with especial attention to topological aspects.

\textbf{2. Moire and Dirac points}

Consider first the Dirac points in TwBLG within the one-electron picture (neglecting correlations).
Two close graphene layers yield  a moire superlattice which
modifies the graphene electron
dispersion and opens gaps both at the primary Dirac point  and the
moire-induced secondary Dirac
point  in the valence band \cite{moire,moire1,1707.09054}.

%\textit{ Ñì. Moir\'e bands in twisted double-layer graphene. R. Bistritzer and A. H. MacDonald \cite{1009.4203}.}

To zeroth order, the low-energy band structure of TwBLG can be considered as two sets
of monolayer graphene Dirac cones (each is four-fold degenerate due to valley and spin) rotated about the $\Gamma$ point by the twist angle. The difference between the  two wave vectors at the point  $K$ (or $K'$) gives rise to the mini
Brillouin zone (MBZ) -- a small hexagon, which is reciprocal to the moire
superlattice \cite{moire}. The Dirac cones near the same valley mix through interlayer hybridization, whereas interactions between distant Dirac cones are  suppressed, so that the valley itself is  a good quantum number.

The  Dirac cones are characterized by a renormalized Fermi velocity $v_F$.
At $v_F \to 0$ there exist three additional Dirac points with opposite winding numbers
($-1$) to the main Dirac point ($+1$). For $v_F = 0$ when all four Dirac points merge,
the winding number is $-2$, since the total winding number cannot change \cite{Goerbig}.
At exactly the first magic angle, the Dirac point at each corner of the MBZ ($K_s$ and $K_{s'}$) becomes a parabolic band touching with winding number $-2$, similar to bilayer graphene with Bernal stacking (except that the two corners have the same winding number) \cite{moire}.

 When crossing the van Hove energy with doping the topology
of the Fermi surface
changes \cite{VH}.  The winding number drops from $-1$ or $+1$ (depending on
the conduction band) to 0
since higher energy contours encircle the  MBZ $\bar{\Gamma}$ symmetry point
where no Berry curvature exists.
Thus we have a topological transition
%from small to large Fermi surface
in the presence of van Hove singularities.

%\textit{Ñì. îáçîð Dirac fermions in condensed matter and beyond.Mark Goerbig and Gilles Montambaux \cite{1410.4098}}

%\textit{Ñì. Charge inversion and topological phase transition at a twist angle induced van Hove singularity of bilayer graphene. Youngwook Kim, Patrick Herlinger, Pilkyung Moon, Mikito Koshino, Takashi Taniguchi, Kenji Watanabe, and Jurgen H. Smet. Nano Lett., 2016, 16 (8), pp 5053--5059. \cite{1605.05475}}

%?? (oboznachenija)There is another lowest energy conduction band which emerges from the ${K}'_t$ and ${K}'_b$ Dirac cones of the single layers. It is mirror symmetric to the band from the ${K}_t$ and ${K}_b$ Dirac cones as seen in the iso-energetic contour plots in Fig. 2. {\it States of both conduction bands do not mix.} They are hereafter referred to as the ${K}$ and ${K}'$ conduction bands.
%On the contour plots the winding number for each class of orbits has been marked.
%The contour at the van Hove energy is plotted in red.

%Arrows along the contour lines indicate whether orbits are executed clock- or anti-clockwise for a chosen perpendicular field orientation.

%\textit{Ñì. \cite{0502139}}
\textbf{3. Mott transition  in TwBLG}

According to the experimental data \cite{moire}, at $|n|> n_s/2$ the Shubnikov-de Haas oscillation frequencies in TwBLG correspond to straight lines which extrapolate to zero  at the half-filling densities. The authors of \cite{moire} suppose that this  may mean small Fermi pockets resulting  from charged quasiparticles near a Mott-like insulator phase, and the halved degeneracy of the pockets might be related to the spin-charge separation in the Mott insulator, similar to the situation in cuprates \cite{Nagaosa}.
In other words, we have a partial Mott transition in the localized spin states subsystem (formation  of the Hubbard subbands violating the Fermi liquid picture). Note that in this sense the narrow band Hubbard ($t-J$) model can be effectively represented as a two-band s-d exchange model \cite{Scr}.

%On a triangular lattice, however, the frustration prevents the fully anti-parallel alignment of adjacent spins. Possible ordering schemes include 120° Néel order and rotational symmetry breaking stripe order. 35 It is yet unclear whether the spin-singlet ground state in TwBLG is fulfilled by any of the above ordering schemes or simply disordered at low temperatures.
%??In the HFIPs of TwBLG it is also possible that the ordering, if any, occurs in conjunction

Thus we have a situation of strong correlations.
As concluded in  \cite{moire}, a theoretical treatment of the TwBLG
problem can be performed within a two-band Hubbard model (including the valley degrees of freedom) on a frustrated triangular lattice.
Then
%In our problem, the basis is changed due to valleys and two layers,
we have to use the SU(4) basis \cite{Balents}. However, these two valleys can be treated as independent in zero order consideration.

%\textit{Ñì. \cite{0901.4103}}

%We start from the insulating spin model in Eq. (2),
%we include the full Hilbert space of the Hubbard model in Eq. (1), and begin with a conventional metallic state with a Fermi surface. The idea is to turn up the value of U at an odd-integer filling of the elctrons so that

To treat the metal-insulator (Mott-Hubbard) transition,  one represents conventionally  the electron annihilation operator as a product of a charged boson $b_i$ and a neutral spinful fermion (spinon) $f_{i\sigma}$, so that in the rotor representation (see \cite{Senthil,Sachdev})
\begin{equation}
\label{1}
c_{i\sigma} = b_i f_{i\sigma}.
\end{equation}
With increasing the Hubbard $U$, the spinless boson system at an odd-integer band filling undergoes  a superfluid-to-Mott insulator transition. In a mean
field description, the spinons are free (non-interacting), despite strong correlations in the electronic system.
If the boson $b_i$ is condensed
$(\langle b \rangle \neq 0)$ we get  the Fermi liquid (FL) phase of
the physical electrons:  when replacing $b$ by its $c$-number average  $\langle b \rangle$, the $f_\sigma$ fermions acquire the same quantum numbers as the initial  electrons, so that the $f_\sigma$ Fermi surface describes a conventional metal.
If the boson is gapped and consequently uncondensed,
a spin liquid Mott insulator occurs, where Fermi surface of neutral fermionic excitations (spinons) survives.
The Mott insulator for the bosons is also an insulator state for the electrons with a gap to all charged excitations, and there is a continuous transition to an insulator with a ``ghost'' spinon Fermi surface.
Thus we have the situation of deconfinement  where  charge and spin degrees of freedom are separated, and the gauge field can play an important role. Away from half-filling, the Bose holon operators should be introduced using other slave particle representations, see \cite{Nagaosa}.

%The phase transition at $g_c$ between the two phases is driven by the condensation of the boson $b_r$.

The flat band situation with large effective mass is somewhat similar to that in heavy fermion (Kondo) systems where f-electron states become delocalized and take part in the Fermi surface even in the absence of ``direct'' hybridization \cite{I17,Sachdev2}.

%The $f_\alpha$ Fermi surface survives in this insulator, and describes a continuum a gapless, neutral spin excitations --- this is the spinon Fermi surface.

\textbf{4. Dirac points and spinons in the strongly correlated case}

The Mott transition on the honeycomb lattice corresponding to grahene (the situation on the bilayer graphene triangular lattice is similar) has been investigated in Refs. \cite{0502139,jafari}. In this case the correlated metallic state is  a semimetal
 containing gapless electronic excitations  at isolated Fermi points in the
Brillouin zone only. These points  are essentially the Fermi surface of electrons.

%Therefore we come to a somewhat different picture in comparison with the one-electron picture considered above in Sect.2.
The  states near the Fermi points have
a  Dirac-like spectrum, and the problem can be analyzed within the corresponding relativistic  formalism. The low energy action for the neutral Dirac spinons $\Psi$ in the insulating
phase has the  structure
\begin{eqnarray}
S&=&\int d^{3}x \sum_\mu \sum_{\sigma=1}^N\bar{\Psi}_{\sigma}
 (\partial_{\mu}-ia_{\mu})\gamma_{\mu}\Psi_{\sigma}
\label{Q}
\end{eqnarray}
%\end{equation}
where integration is performed in 2+1 dimensions,  $\gamma_\mu$ are the Dirac matrices, $\bar{\Psi}_{\sigma} \equiv \Psi^{\dag}_{\sigma} \gamma_{0}$, and $a_\mu$ is an emergent gauge field associated with the spinon-boson decomposition of the electron operator (1).
Note that the Dirac excitation spectrum can be formed even if the bare lattice electron dispersion does not lead to such  a  spectrum (e.g., for the square
and kagome lattices) \cite{Sachdev}.
% by allowing for non-zero average aµ fluxes on the plaquettes, and optimizing these fluxes variationally, it is found that the resulting 'flux' states do often acquire a Dirac excitation spectrum.

%Despite strong correlations in the electronic system, spinons can be considered as free particles.
%gauge redundancy introduced by the decomposition in Eq. (12)??.

Depending upon the details of the lattice,  $a_\mu$ can be a U(1) or
SU(2) gauge field. For a large number of flavors $N$ (determined by the number of the Dirac points in the Brillouin zone), the action $S_D$ describes a conformal field theory (CFT). Thus we have a scale-invariant, strongly interacting quantum state with a power-law spectrum for all excitations, well-defined quasiparticles being absent.
This state is labeled as an algebraic spin liquid (see  \cite{Sachdev,Nagaosa}).
%Unlike quantum electrodynamics (U(1) symmetry) and quantum chromodynamics (SU(3) symmetry) we have SU(4) situation. The problem is similar to Yang-Mills theory where self-interaction of gauge fields occurs, as well as in quantum chromodynamics.  Gluons (Yang-Mills gauge bosons) do not screen gauge charges, but lead to asymptotic freedom and confinement.
Note that one of the ways to obtain a deconfined phase  is to include gapless excitations which carry gauge charges.  These  excitations can screen the gauge interaction to make it less confining  \cite{Nagaosa}.

Unlike true deconfined phases where noninteracting
quasiparticles become free at low energies, here
deconfinement  means only that the gapless charged
particles remain gapless, but are not quite free.  The
corresponding gapless spin liquids are obtained from the
staggered flux liquid (sfL)  and uniform RVB (uRVB) phases.
%[given by Eq. (115) with $\Delta = 0$]belong to this case.
 The uRVB state  leads upon doping to strange-metal states
with a large Fermi surface, so that the picture of isolated spinon
Dirac points somewhat changes \cite{Nagaosa}.

The FL phase contains boson condensation which restores the quasiparticle picture.
Therefore the low-energy excitations in the FL phase are
described by electron-like quasiparticles and this phase corresponds to a FL phase
of electrons.

The dynamics for the $U(1)$ gauge field arises  owing to  screening by
bosons and fermions, both  carrying gauge charge. In the low doping
case one can  take into account screening by fermions only.
After integrating out $\Psi$ in (\ref{Q}) the  effective
partition function for the U(1) gauge field reads \cite{Kim}
\begin{eqnarray}
\cal Z&=&\int Da_{\mu}\exp\Big( -\frac{1}{2}\int\frac{d^3q}{(2\pi)^3}a_{\mu}
({q})\Pi_{\mu\nu}a_{\nu}(-{q})\Big), \nonumber \\
\Pi_{\mu\nu}&=&\frac{N}{8}\sqrt{{q}^2}\Big(\delta_{\mu\nu}
- \frac{q_{\mu}q_{\nu}}{{q}^{2}}\Big).
\label{Pi}
\end{eqnarray}
The polarizability $\Pi$ makes the
gauge coupling $a_\mu j^\mu$ a marginal perturbation at the free fermion fixed
point.
%Importantly however we should note that

Consider the electron Green's function.  In the leading order in $1/N$, it was found that
\begin{equation}
 G(x)=\langle b^\dagger( x) b(0)\rangle_0
\langle f( x) f^\dagger(0) \exp (i\int_0^{x} dx \cdot a) \rangle
\end{equation}
where $\int_0^{x}d x$ is the integration along the straight return path
and $\langle...\rangle$ stands for integrating out the gauge fluctuations \cite{Nagaosa}.
Then one obtains
\begin{equation}
 G(x)
\propto (x^2)^{-(2-\alpha)/2}
\label{b}
\end{equation}
with the exponent $\alpha \sim 1/N$ being the anomalous dimension; for the square-lattice antiferromagnet $\alpha = 32/(3 \pi^2 N)$.
These results describe a partial confinement of spinons and bosons coupled by the gauge field.
The conductivity is determined by the contributions of both fermions and bosons which cannot be considered as independent quasiparticles \cite{Ioffe,Nagaosa}.

\textbf{5. Topology of Lifshitz transitions}

%\textit{\cite{0601372,1701.06435}.}

The electron states in a strongly correlated system need not to have purely quasiparticle nature. They can be described by both poles and branch cuts of the Green's function, cf. Eq. (\ref{b}). If the suppression of the quasiparticle residue $Z$  is strong,   the
pole in the Green's function can be even transformed to the zero,
%which corresponds to the special case of $\gamma = 1$,
$G(E) \propto E + \varepsilon(\bf{k})$, which means formation of the energy gap and takes place, e.g., for the Mott transition \cite{Volovik}.
The violation of the standard Fermi-liquid picture can be described in terms
of the formation of the Luttinger surface which is the surface of zeros
of the electron Green's function \cite{YRZ}.

The Lifshitz transitions (in particular, those discussed in Sect. 2) can be viewed as quantum phase transitions  with  change of the topology of the Fermi surface (FS), but without symmetry breaking. The topology of FS is characterized not only by its shape. FS itself  is  the singularity in the Green's
function, which is topologically protected: it is the vortex
line in the  frequency-momentum space \cite{Volovik,Volovik1}.
%his situation takes place even
Formally, in the Mott insulating phase FS does not exist.
However,  the topology of FS is preserved if we take into account the Luttinger contribution.
Then the Luttinger theorem (the conservation of the volume enclosed by FS independently of the interaction strength) is still valid \cite{Volovik,Sachdev2}.
It should be noted that the Fermi surface combined from the poles and zeros  is a whole object which cannot have holes and edges  \cite{Volovik}. A similar picture occurs in cuprates where the Fermi points are stretched into arcs to form a large closed Fermi surface \cite{Nagaosa}.

As for the non-Fermi liquid behavior, the flat band can be also treated as the  Khodel-Shaginyan fermion condensate caused by electron-electron interaction \cite{Khodel}, where all the states have zero energy.

%However, this quantum phase transition is not topological since the topological invariant does not change across the transition.

The Lifshitz transition for  bilayer graphene is governed by the conservation
of the topological charge $N_2$. Merging
of the two conical points with $N_2 = 1$ leads to formation
of the Dirac node with quadratic dispersion and the topological charge $N_2 = 2$.
Interaction between the layers may
lead to several possible scenarios of the geometry of the fermionic spectrum in bilayer
graphene.
In particular, the $N_2 = 2$ Dirac point can split into four Dirac conical points with
$N_2 = \pm 1$ (trigonal warping). The total topological charge is conserved,
$N_2 = 1 + 1 + 1 - 1 = 2$  \cite{Volovik}.

Thus the topological treatment may provide an interpolation from the one-electron Dirac points to the Dirac fermions (spinons) at the Fermi points in the strongly correlated Hubbard ($t-J$) model.
%{\it Çàìå÷àíèå: À åñòü ëè òðèãîíàëüíàÿ äåôîðìàöèÿ â äâóñëîéíîì ãðàôåíå ïðè ïîâîðîòå âåðõíåãî ñëîÿ? Áûëî ëè ñëèÿíèå äâóõ äèðàêîâñêèõ êîíè÷åñêèõ òî÷åê?}

\textbf{6. Superconductivity}

Many features of TwBLG
are  similar to those of the cuprate high-$T_c$ materials where  superconductivity  occurs in a close vicinity of  Mott insulator state after passing small antiferromagnetic region with doping. Here the transition may occur through complicated states, including
incommensurate charge and magnetic order, stripes and magnetic phase separation.
 Doping is supposed to frustrate the ground state Neel order so
that the system is pushed across the  transition where the
Neel order is lost and a spin liquid state arises \cite{Nagaosa}.
%However, in this direct approach the connection with superconductivity is not at all clear. Instead it is conceptually useful
Thus the transition to the superconducting state goes continuously via quantum critical point, a pseudogap state forming at finite temperatures in the quantum critical regime. The critical point can have a deconfinement nature.
%one uses  an indirect path starting from a spin-liquid state.
At the same time, in TwBLG the situation seems to be even more clear  since antiferromagnetism is totally suppressed by frustrations of triangular lattice.

Introducing exchange interactions in the narrow band system enables one to treat exotic topological phases and superconductivity \cite{Nagaosa}.
In particular, the d-wave superconducting phase contains both the boson
and fermion-pair condensate.

The superconducting properties can change in the deconfinement situation.
The conventional and strongly correlated superconductors (being topologically ordered states) can be distinguished by flux quantum which equals $hc/(2e)$ in the former case (fermion pairing) and $hc/e$ in the latter case (Bose condensation in the spin-gap state) \cite{Nagaosa}.

Recently, the value $hc/(4e)$ was found in Ref. \cite{Balents}. These authors proposed topologically protected gapless edge states and half-vortices carrying half the usual superconducting flux quantum (effective $4e$ charge superconductor).
 The flat band superconductivity has been  also discussed in Ref.\cite{Volovik2}.

\textbf{7. Conclusions}

According to estimations in Ref. \cite{moire}, we have in TwBLG strong coupling or intermediate coupling situation (the Hubbard $U$ is larger or of order of effective electron bandwidth).
We have considered above the strong correlation limit in TwBLG system in terms of spinon-boson denconfinement.
%Pockets?
At the same time, occurrence of spinons and fractionalized Fermi liquid (FL$^*$) state with non-Fermi-liquid features can take place also in the intermediate coupling case described by the spin-fermion model with suppressed magnetic ordering  \cite{Punk1,Scr}.
The transition from small to large Fermi surface can be connected with the change in statistics of spinons \cite{Punk1}.

With increase of doping, the Fermi level crosses the van Hove singularity in the nearly flat band \cite{VH} and we come to a new strongly correlated state.
A similar situation in cuprates (pinning of the Fermi level to the van Hove singularity and the formation of flat bands in the two-dimensional $t-t'$ Hubbard model) was considered in Ref.\cite{Katanin}.
The corresponding theoretical treatment for TwBLG requires further investigations.

The research was carried out within the state assignment of FASO of Russia (theme ``Flux'' No AAAA-A18-118020190112-8 and theme ``Quantum'' No. AAAA-A18-118020190095-4) and supported in part by the Russian Foundation for Basic Research (project no.
16-02-00995).

\end{document}